# Structural phase transitions and magnetic superexchange in MIAgIIF3 perovskites at high pressure

Łukasz Wolański[a], Marvin Metzelaars,[b] Jan van Leusen,[b] Paul Kögerler[b][c]* and Wojciech Grochala[a]*

*This work commemorates the 200th anniversary of the birth of Louis Pasteur and the 100th anniversary of the birth of Rudolf Hoppe*

[a]    Dr. Ł. Wolański, Prof. W. Grochala
Centre of New Technologies, University of Warsaw, S. Banacha 2c, Warsaw 02-097, Poland
[b]    Dr. M. Metzelaars, Dr. J. van Leusen, Prof. P. Kögerler
Institute of Inorganic Chemistry, RWTH Aachen University, Aachen 52074, Germany
[c]    P. Kögerler
Peter Grünberg Institute (PGI-6), Forschungszentrum Jülich GmbH, Jülich 52425, Germany

*   e-mail: paul.koegerler@ac.rwth-aachen.de, w.grochala@cent.uw.edu.pl

Supporting information for this article is given at the end of the document.

**Abstract:** Pressure-induced phase transitions of MIAgIIF3 perovskites (M = K, Rb, Cs) have been predicted theoretically for the first time for pressures up to 100 GPa. The sequence of phase transitions for M = K and Rb consists of an orthorhombic to monoclinic and back to orthorhombic transition, associated with progressive bending of infinite chains of corner-sharing [AgF6]4– octahedra and their mutual approaching via secondary Ag⋯F contacts. In stark contrast, only a single phase transition (tetragonal → triclinic) is predicted for CsAgF3, associated with substantial deformation of the Jahn Teller-distorted first coordination sphere of Ag(II) and association of the infinite [AgF6]4– chains into a polymeric sublattice. The phase transitions markedly decrease the coupling strength of intra-chain antiferromagnetic superexchange in MAgF3 hosts lattices.

**Introduction**
There has been a great upsurge of interest in halide perovskites, particularly due to their ability to serve as photochemically active materials in solar cells.1,2 Among all halides, fluorides tend to be the least studied, since they usually exhibit very large fundamental band gaps, and the corresponding perovskites are poor electric conductors. Divalent silver ($Ag^{2+}$) fluorides stand out in this group due to their moderate band gaps approaching the UV/vis edge,3 and substantial involvement of F(2p) orbitals in covalent chemical bonding to Ag.4,5,6 These features also permit the markedly strong superexchange between spin-½ 4d9 metal centers mediated by fluoride bridges.7,8 The structures and selected properties of MAgF3 perovskites (M = K, Rb, Cs), have been studied by Hoppe et al. half a century ago,9 and they have been critically scrutinized more recently by one of our groups.10,11,12,13 It turns out that dark brown MAgF3 perovskites (M = K, Rb, Cs) may serve as hosts of exceptionally strong antiferromagnetic superexchange characterized by superexchange constants, $J_{1D}$, ranging from ca. –100 meV (~ –800 cm–1) for K+ to –180 meV (~ –1450 cm–1) for the Cs+ analogue.8,11 While the ambient-pressure crystal structures of the MAgF3 series (M = K, Rb, Cs) are known, their polymorphism is a virtually uncharted area, in contrast to the pressure-induced phase transitions for their parent compound, AgF2.14,15

Therefore, we here have explored theoretically the impact of high pressures (up to 100 GPa) on the crystal structures as well as on the magnetic properties of the MAgF3 series. By comparing results obtained for M = K, Rb and Cs, we aim at understanding the trends governing their structural preferences and magnetic characteristics.

**Results and Discussion**

Our approach is based on solid-state Density Functional Theory (DFT) computations carried out with the help of self-learning algorithms (XtalOpt r11.0)16,17 (cf. Computational Details section), while accounting for ferromagnetic and diverse antiferromagnetic models (cf. Supplementary Information, SI).

Calculations carried out for p = 0 GPa and T = 0 K result in delivered ground state structures in agreement with experiment9,10,11 and with a recent computational study.12,13 Both the K and Rb salts crystallize in orthorhombic Pnma cells (Fig. 1) (distorted GdFeO3-type perovskite) while the Cs analogue adopts a higher symmetry tetragonal I4/mcm cell (also representing a distorted perovskite). The infinite AgF3– chains are bent for the K salt, and are ideally linear for the Cs salt. The Rb structure represents an intermediate between these two structures. The AgF3– chains are characterized by strong antiferromagnetic superexchange, and thus the ranking of J1D values follows closely the Ag–F–Ag angle. Due to marked similarity of the Pnma and I4/mcm cells these polymorphs are labelled here jointly as structure A (Fig.1).

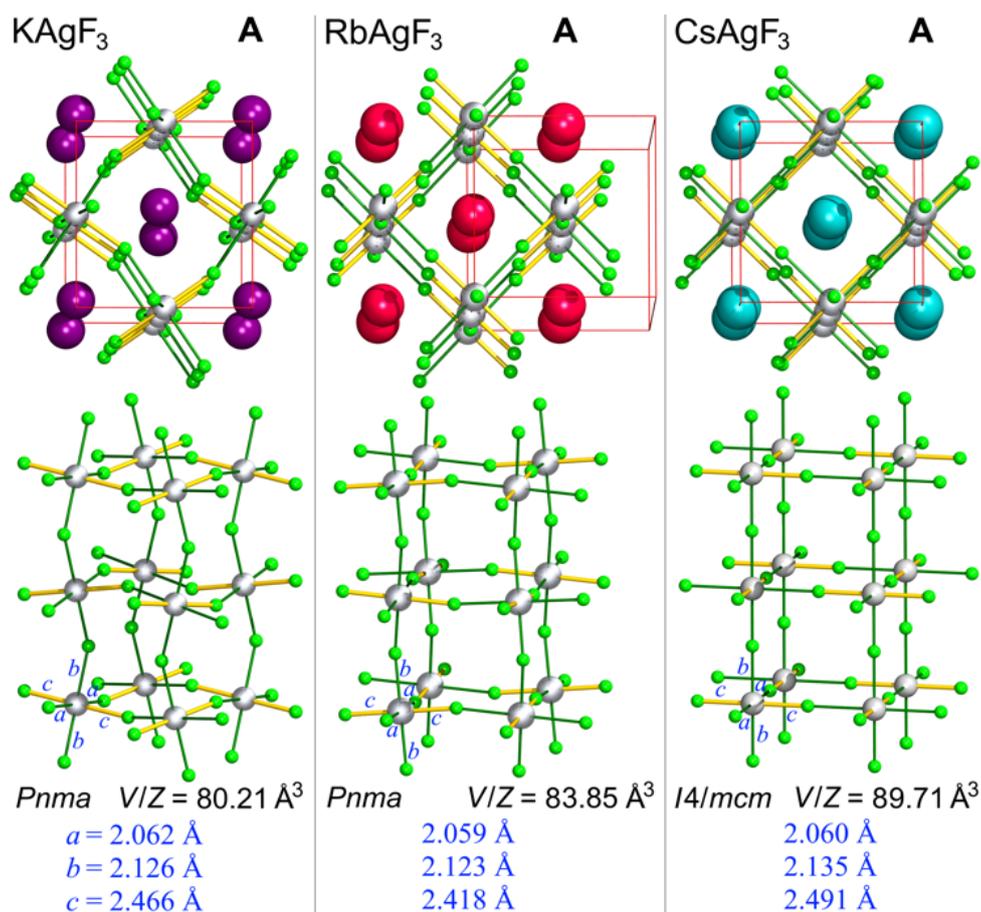

Figure 1. Minimum-enthalpy equilibrium structures of the A isomers of MAgF3 (M = K, Rb and Cs). Only Ag–F bonds are shown for clarity, and M+ cations are omitted in the bottom figures. Unit cell edges are emphasized as red lines. Lengths of Ag–F bonds (a, b, c) are indicated by different colours (green: short, yellow: longer).

Our calculations reveal the presence of two consecutive phase transitions for KAgF3 and RbAgF3 but only one for CsAgF3 (Fig. 2). Importantly, the polymorphs of K and Rb derivatives are identical; we denote them here as B and C, and their crystal structures are shown in Fig. 3. On the other hand, an entirely different form labelled D is the sole high-pressure polymorph of CsAgF3 up to 100 GPa (Figs. 2 and 3). As it could be expected, the pressure needed for the first phase transition to occur varies from ca. 4.5 GPa for K, via ca. 12.5 GPa for Rb up to ca. 35 GPa for Cs analogue (Fig. 2). On the other hand, the next phase transition for K salt is expected at 38 GPa, for Rb one at 64 GPa, while it is absent for Cs up to 100 GPa. Such behavior shows a clear and monotonic trend in the series of alkali metal cations with strongly increasing ionic radius.

B is monoclinic, P21/m, while C is orthorhombic and it formally belongs to the same space group as the ground state structure A (Pnma) for K and Rb derivatives. Yet, there are noticeable differences between them (Fig. 3). For example, computation for form A of KAgF3 at 10 GPa reveals the following coordination sphere of Ag: 2× 2.09 Å (for Ag–F bonds parallel to the propagation direction of the infinite chains), and 2× 2.03 Å plus 2× 2.30 Å (yielding an antiferrodistortive bond pattern within [AgF2] sheets). Simultaneously, form B of this compound at 10 GPa features the following Ag–F bonds: 2× 2.08 Å (for Ag–F bonds parallel to the propagation direction of the infinite chains), and 2x 2.04 Å plus 2× 2.40 Å (within [AgF2] sheets). Clearly, the longest Ag–F distance in the first coordination sphere of Ag is less squeezed in B than in A, resulting in more favorable enthalpy for the former at 10 GPa.

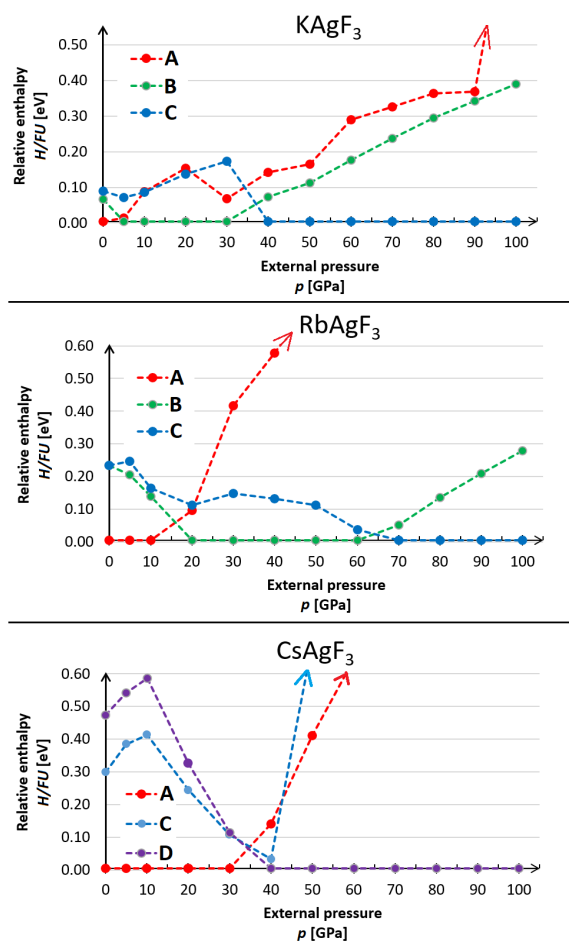

Figure 2. Influence of external pressure on minimum-enthalpy equilibrium structures of MAgF3. For each member of the series the relative enthalpies per formula unit, H/FU, of the relevant polymorphs are shown. Enthalpies of the most stable form for each external pressure define the energy origin.

In this way, the Jahn-Teller distortion of the AgF6 octahedron may be more pronounced,[18,19] and this is associated with a smaller local strain than for the structure A. There are also differences in the first coordination sphere of the alkali atoms in each forms. For example, in the case of KAgF3 at 10 GPa, K+ is surrounded by nine fluoride ligands in structure A (at 2.51 Å, 2× 2.52 Å, 2.61 Å, 2× 2.69 Å, 2× 2.72 Å, and 2.85 Å) with an average bond length of 2.65 Å. Simultaneously, in the polytype B the coordination sphere of K consists of eight bonds: 2.53 Å, 2× 2.53 Å, 2.58 Å, 2.60 Å, 2× 2.62 Å, and 2.97 Å, with an average bond length of 2.62 Å. Thus, the expected increase of the coordination number of alkali metal at elevated pressure[20] is not observed and as such cannot drive the A → B structural transition. Similar conclusions related to local Jahn-Teller effect and to coordination sphere of alkali metal cation may be reached for the respective A → B phase transition for the RbAgF3 homologue. The AgF6 octahedron at 20 GPa exhibits the following bond lengths: 2× 2.01 Å, 2× 2.02 Å, and 2× 2.21 Å in structure A, and 2× 2.02 Å, 2× 2.09 Å and 2× 2.56 Å in structure B. Simultaneously, the coordination number of Rb drops from 11 in A to 10 in B. Again, this is not the change of the coordination sphere of alkali metal cation, but rather the structural flexibility of structure B in releasing the strain associated with squeezing of the AgF6 octahedra, which is beyond the A → B phase transition.

And what are the structural changes associated with the second structural phase transition, B → C? Inspection of the crystal structure of KAgF3 at 40 GPa in structure B and C leads to the following conclusions: the bonding pattern in the first coordination sphere of Ag changes from a more regular one resembling 4+4 coordination, 2× 2.00 Å, 2× 2.06 Å plus 2× 2.43 Å and 2× 2.44 Å, to a highly irregular one 2.00 Å, 2.05 Å, 2.09 Å, 2.11 Å, 2.28 Å, plus 2.44 Å and 2.46 Å, resembling (4+1)+2 coordination. In other words, AgF4 squares with four secondary contacts are substituted by irregular AgF5 pyramids with two additional longer interactions. A similar mechanism has been seen to drive the pressure-induced phase transitions in binary AgF2, where quasi-pentacoordinated units were detected.[14,15]

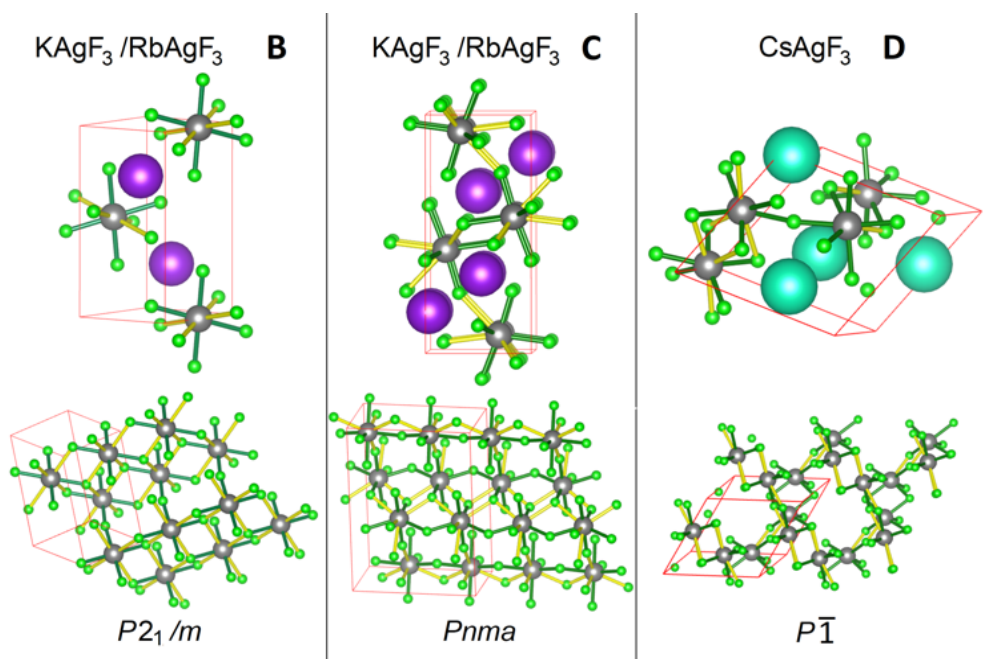

Figure 3. Minimum-enthalpy equilibrium structures of the B polymorphs of KAgF3/RbAgF3 under external pressure of 30 GPa (left column), C polymorphs of KAgF3/RbAgF3 (middle column) and the D form of CsAgF3 at 100 GPa (right column). The Ag–F bonds are shown in green (short bonds) and yellow (longer bonds). In some cases, M+ cations are omitted for clarity. Unit cells have been emphasized by solid lines.

The sole pressure-induced phase transition predicted for CsAgF3 salt is also associated with large changes of the first coordination sphere of Ag(II). The parent structure at 30 GPa features the following Ag–F bond lengths: 2× 1.98 Å, 2× 1.99 Å and 2× 2.11 Å. The post-perovskite structure D at 40 GPa shows two independent Ag sites, both with highly irregular Ag2+ coordination (one pentacoordinated: 1.97 Å, 2.04 Å, 2.10 Å, 2.13 Å, 2.34 Å, and another hexacoordinated: 1.99 Å, 2.02 Å, 2.13 Å, 2.14 Å, 2.23 Å, 2.32 Å). Here, the former AgF3– chains polymerize into a complex 3D network. As in the case of K and Rb salts, the alkali metal cation experiences a counterintuitive decrease of the coordination number upon transition, with 12-coordinated Cs in the ambient-pressure tetragonal structure, and 11-coordinated Cs in the high-pressure structure.

The described structural transitions heavily impact magnetic properties of MAgF3 perovskites. All ambient-pressure structures comprise linear or only slightly bent infinite AgF3– chains, which produce strong 1D antiferromagnetic superexchange.10,11 The calculated J1D values range from ca. –100 meV for the K salt, to –180 meV for the Cs analogue.11 However, the A → B structural phase transition leads to bending of the Ag–F–Ag angle from 161° to 152° (for the K salt at the transition), and from 143° to 117° (for the Rb salt at the transition). Analogous A → D transition for Cs salt is also associated with remarkable reduction in the Ag–F–Ag angle from 180° to 159°. Not surprisingly, this implies that antiferromagnetic intra-chain superexchange energy must weaken substantially down to ca. –50 meV for K, and only –5 meV for Cs. Deformation of the Ag–F–Ag angle in the case of the Rb salt at 20 GPa is so pronounced (117°) that a weak ferromagnetic superexchange predominates (ca. +5 meV).

**Conclusions**

Structural phase transitions for MAgF3 perovskites (M = K, Rb, Cs) have been predicted theoretically for pressures up to 100 GPa, exhibiting similarities between KAgF3 and RbAgF3, from which CsAgF3 differs starkly. Surprisingly, a decrease of the coordination number of alkali metal is observed as pressure increases, and the critical pressure of the first transition rises together with the ionic radius of the alkali metal,21 contrary to what is usually observed.22 Release of the strain here is associated with compression of the Jahn-Teller-distorted [AgF6]4– octahedron present in the low-pressure structures. This Jahn Teller distortion of the 4d9 Ag2+ ions is key to the symmetry lowering in the high-pressure polymorphs of all title compounds. While the second phase transition for K and Rb salts is associated with a formal increase of symmetry from monoclinic to orthorhombic, yet in contrast to structure A, in structure C Ag does not sit at the special position and its local coordination sphere is highly irregular.

The behavior of KAgF3 strongly contrasts with that of KCuF3 relative23,24 (which is actually isostructural with CsAgF3) where pressures as low as 8 GPa lead to significant reduction of the Jahn-Teller distortion without leading to any structural phase transition.25 Very different behavior of CuII and AgII salts may be understood in terms of the much stronger covalence of the metal-fluorine bonds in the latter.4,5 On the other hand, the behavior of CsAgF3 seems to be unprecedented among fluoride perovskites. While triclinic post-perovskites are known for AgICuIIF3 and NaICuIIF3,26 they lack the presence of a strongly polymerized MF3– anionic sublattice, such as the one seen in structure D of CsAgF3.

For all members of the MAgF3 series studied here, phase transitions impact the Ag–F–Ag angle that substantially decreases upon the phase transitions, and – in agreement with the Goodenough-Kanamori rules – this leads to substantial (from 2- to 40-fold) weakening of the antiferromagnetic superexchange energy for K and Cs salts at the transition; in the case of Rb salt at 20 GPa the Ag–F–Ag angle already is close to perpendicular so that weak ferromagnetism is predicted.

## Computational Details

Solid-state Density Functional Theory (DFT) structural screening was carried out with using self-learning algorithms implemented in XtalOpt r11.0.[16,17] Aside from random structure generation, several structure types were manually fed into the structure pool. These were the available experimental structures and models prepared by Ag+ → K+ substitutions of $Ag_2F_3$ and $Ag_3F_4$ high-pressure structures[27] (note that the ionic radii of Ag+ and K+ are quite similar). DFT calculations utilized the Perdew-Burke-Ernzerhof exchange-correlation functional,[28,29] and the projector-augmented-wave method[30] with appropriate pseudopotentials[31] from v.54 dataset as implemented in the VASP 5.4.4 code.[32] The cut-off energy of the plane wave basis set was equal to 950 eV with a self-consistent-field convergence criterion of $10^{-6}$ eV. The following number of formula units in the unit cell were tested: 2, 3, 4 and 6. Numerous equilibrium structures were obtained and symmetrized. In the second stage, the most promising (lowest enthalpy) structures obtained from preliminary screening were optimized in various ferromagnetic and antiferromagnetic models using the DFT+U method with U = 5.5 eV and J = 1.0 eV.[33,34] Perdew-Burke-Ernzerhof exchange-correlation functional revised for solids (PBEsol)[35] functional was used with tight convergence criteria. For all structures their enthalpies were compared at each pressure to determine the minimum-enthalpy structures.
In the final stage, to estimate the strength of magnetic superexchange, for each investigated external pressure one minimum-enthalpy structure was chosen and single-point energy calculations for different magnetic configurations were performed at the DFT+U (PBEsol) level of theory.
All figures of structures were prepared using the VESTA software.[35]

## Acknowledgements

The research was supported by Polish National Science Center (NCN) and Deutsche Forschungsgemeinschaft (DFG) within BEETHOVEN2 project (2016/23/G/ST5/04320, KO 3990/8-1). Computations were carried out with the support of the Interdisciplinary Centre for Mathematical and Computational Modelling (ICM), University of Warsaw under grants no. G49-17 and GA83-34.

Keywords: fluorides • silver • high pressure • magnetic superexchange • phase transitions • perovskites

# SUPPORTING INFORMATION

# Structural phase transitions and magnetic superexchange inter in M$^I$Ag$^{II}$F$_3$ perovskites at high pressure

Łukasz Wolański[a], Marvin Metzelaars,[b] Jan van Leusen,[b] Paul Kögerler[b][c]* and Wojciech Grochala[a]*

**Contents:**

S1. Investigated magnetic variants of KAgF$_3$ **A** structures

S2. Investigated magnetic variants of KAgF$_3$ **B** structures

S3. Investigated magnetic variants of KAgF$_3$ **C** structures

S4. Investigated magnetic variants of RbAgF$_3$ **A** structures

S5. Investigated magnetic variants of RbAgF$_3$ **B** structures

S6. Investigated magnetic variants of RbAgF$_3$ **C** structures

S7. Investigated magnetic variants of CsAgF$_3$ **A** structures

S8. Investigated magnetic variants of CsAgF$_3$ **D** structures

Table S1. Relative enthalpies of investigated magnetic models of chosen **KAgF$_3$** structures.

Table S2. Relative enthalpies of investigated magnetic models of chosen **RbAgF$_3$** structures.

Table S3. Relative enthalpies of investigated magnetic models of chosen **CsAgF$_3$** structures.

S9. Selected crystallographic .cif files from DFT calculations.

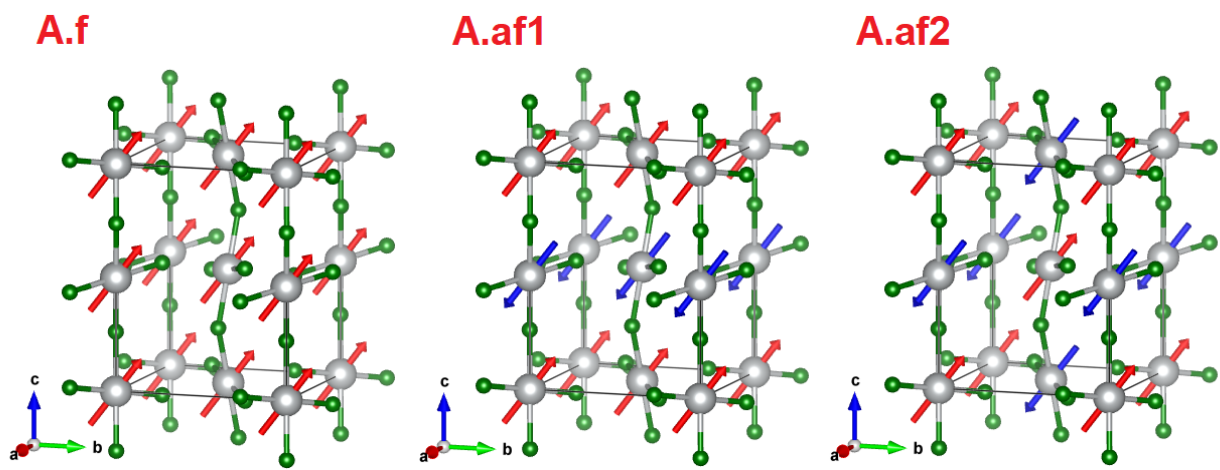

**Figure S1.** Investigated magnetic variants of **KAgF$_3$** structures of type **A**, with ferromagnetic (**A.f**) and antiferromagnetic (**A.af1** and **A.af2**) spin lattices. Only the AgF$_3^-$ sublattices are shown. Ag–F bonds are indicated for interatomic distances not exceeding 2.2 Å. All visualized example structures are for $p$ = 0 GPa.

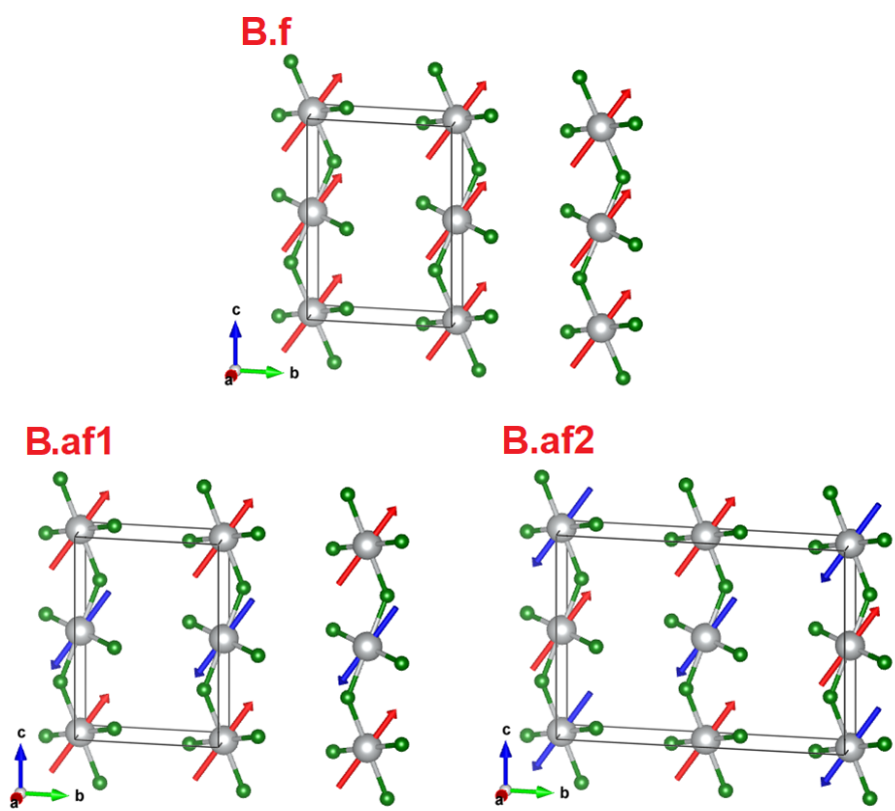

**Figure S2.** Investigated magnetic variants of **KAgF$_3$** structures of type **B**, with ferromagnetic (**B.f**) and antiferromagnetic (**B.af1** and **B.af2**) spin lattices. Only the AgF$_3^-$ sublattices are shown. Ag–F bonds are indicated for interatomic distances not exceeding 2.2 Å. All visualized example structures are for $p$ = 0 GPa.

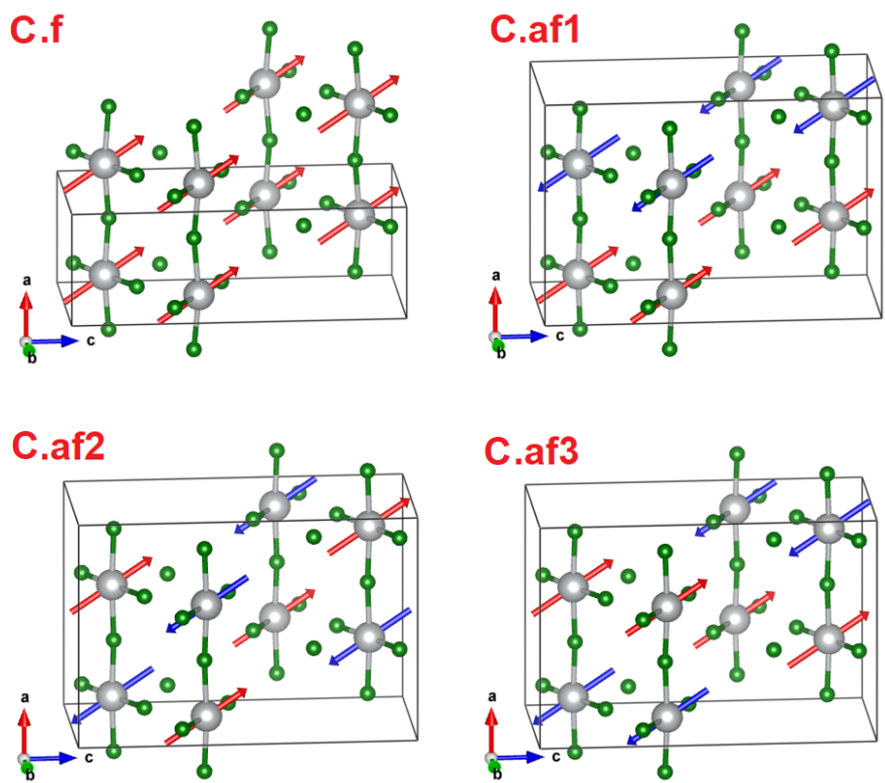

**Figure S3.** Investigated magnetic variants of **KAgF₃** structures of type **C**, with ferromagnetic (**C.f**) and antiferromagnetic (**C.af1**, **C.af2** and **C.af3**) spin lattices. Only the AgF$_3^-$ sublattices are shown. Ag–F bonds are indicated for interatomic distances not longer than 2.2 Å. All visualized example structures are for $p$ = 0 GPa.

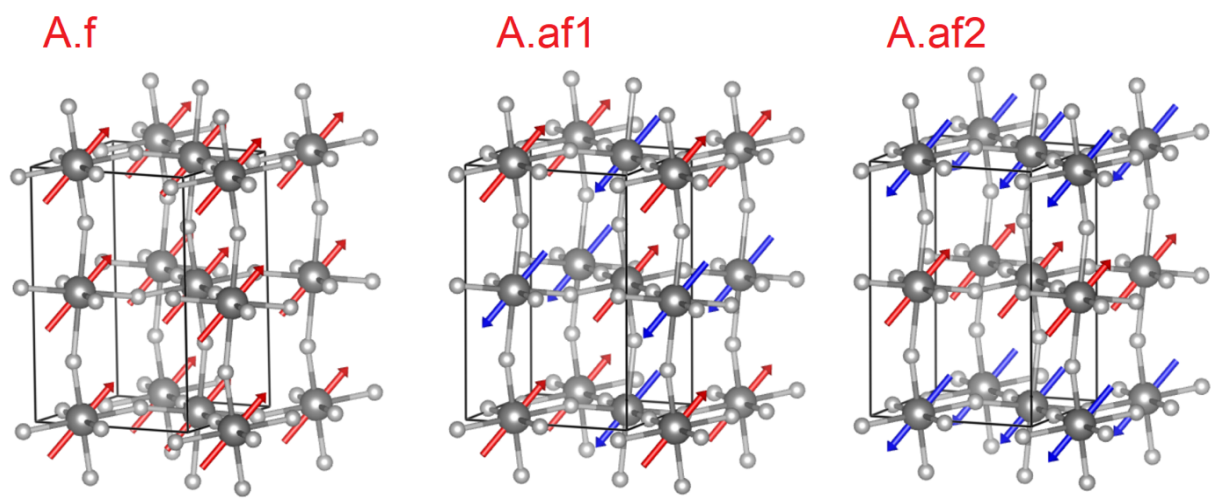

**Figure S4.** Investigated magnetic variants of **RbAgF₃** A-type structures for ferromagnetic (**A.f**) and antiferromagnetic (**A.af1** and **A.af2**) spin ordering. Only the AgF$_3^-$ sublattices are shown. All visualized example structures are for $p$ = 0 GPa.

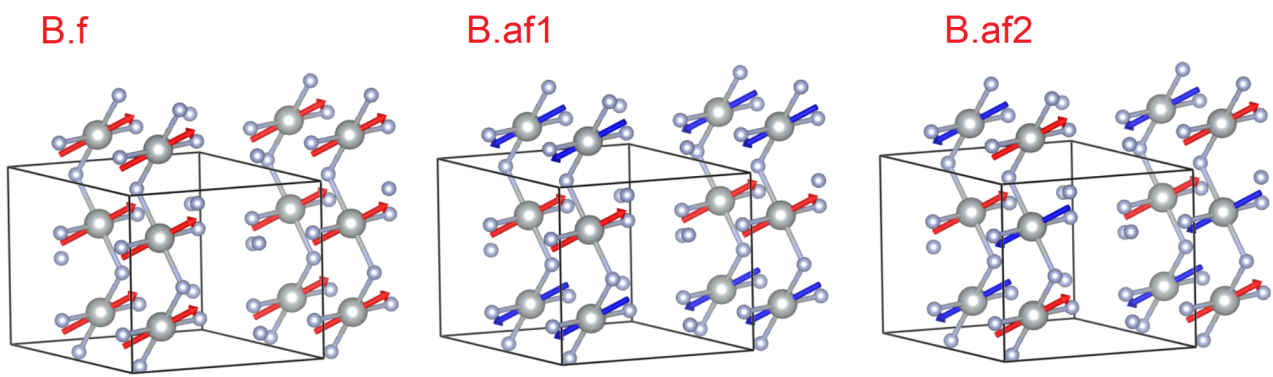

**Figure S5.** Investigated magnetic variants of **RbAgF₃** B-type structures for ferromagnetic (**B.f**) and antiferromagnetic (**B.af1** and **B.af2**) spin ordering. Only the AgF$_3^-$ sublattices are shown. All visualized example structures are for $p$ = 0 GPa.

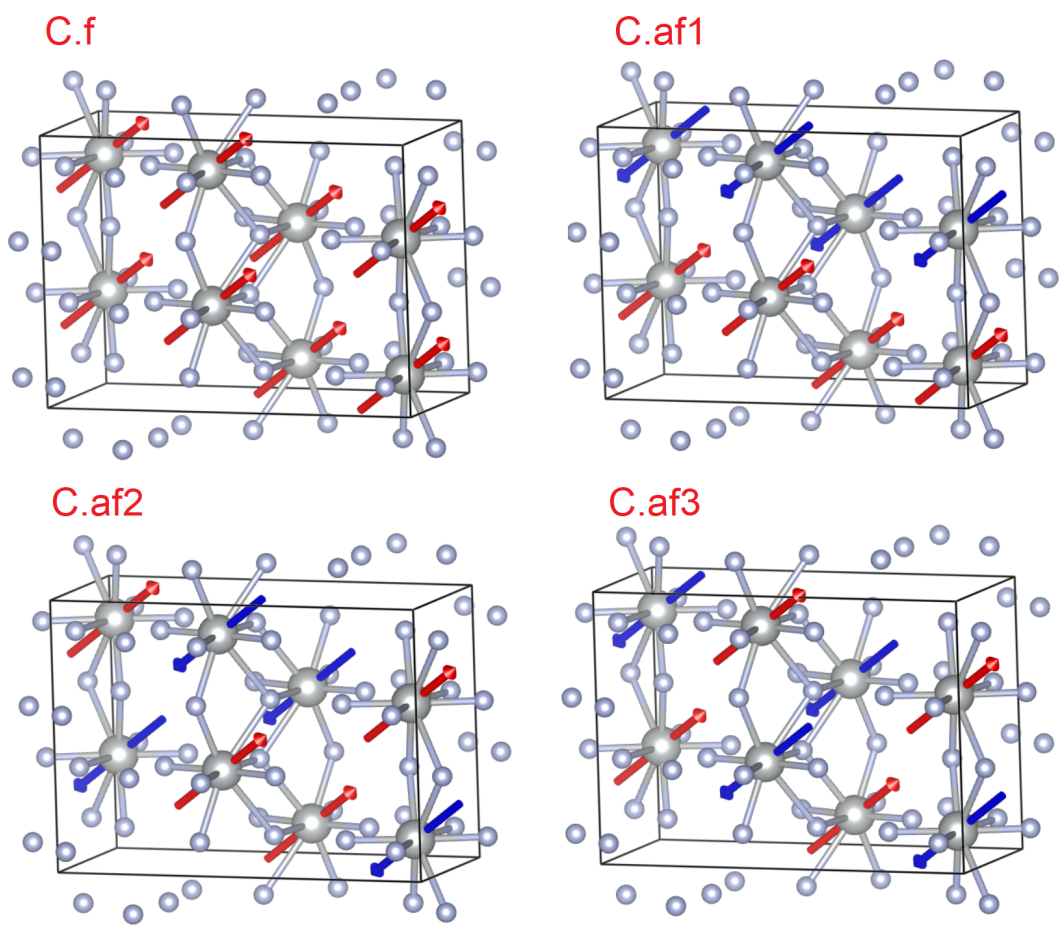

**Figure S6.** Investigated magnetic variants of **RbAgF₃** C-type structures for ferromagnetic (**C.f**) and antiferromagnetic (**C.af1**, **C.af2** and **C.af3**) spin ordering. Only the AgF$_3^-$ sublattices are shown. All visualized example structures are for $p$ = 0 GPa.

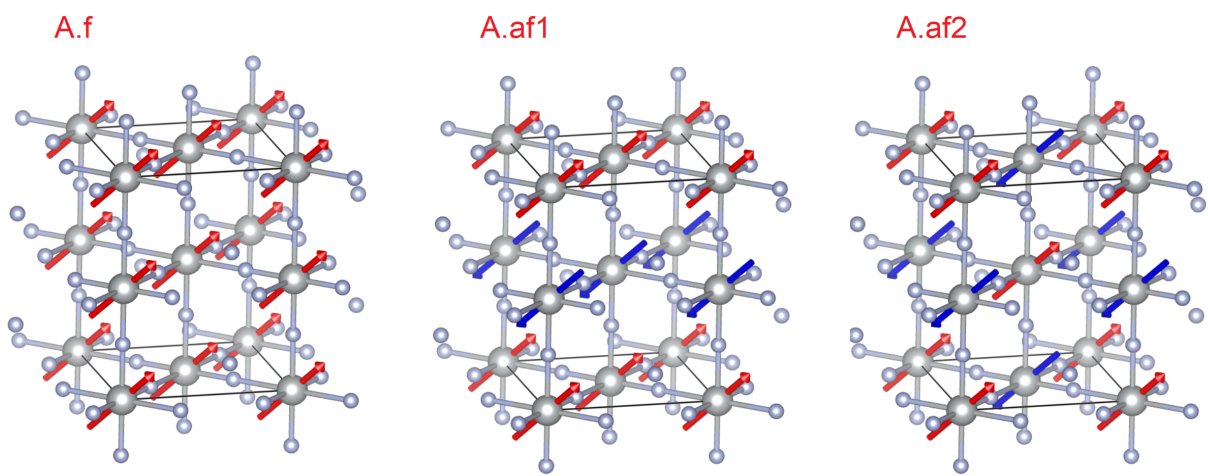

**Figure S7.** Investigated magnetic variants of **CsAgF$_3$** **A**-type structures for ferromagnetic (**A.f**) and antiferromagnetic (**A.af1** and **A.af2**) spin ordering. Only the AgF$_3^-$ sublattices are shown. All visualized example structures are for $p$ = 0 GPa.

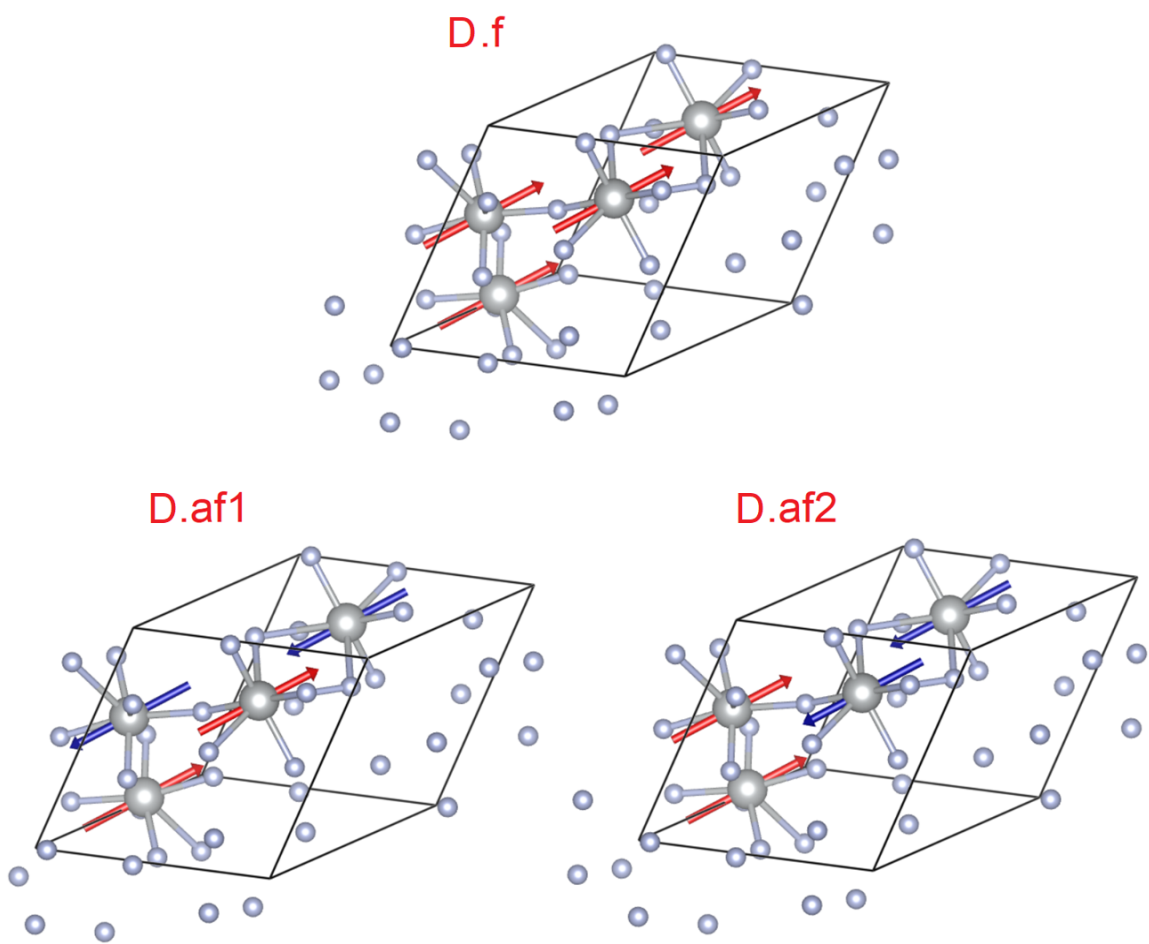

**Figure S8.** Investigated magnetic variants of **CsAgF$_3$** **D**-type structures with ferromagnetic (**D.f**) and antiferromagnetic (**D.af1** and **D.af2**) spin ordering. Only the AgF$_3^-$ sublattices are shown. All visualized example structures are for $p$ = 0 GPa.

**Table S1.** Relative enthalpies (*H/Z* **rel.**) of investigated magnetic models of chosen **KAgF$_3$** structures optimized at DFT+U(PBEsol) level of theory under external pressures ***p*** from the range from 0 to 100 GPa. To facilitate data analysis data are shown as *H/Z* ratio (examined unit cells were of a different *Z* parameter values). Relative enthalpies *H/Z* **rel.** are given with the assumption that the lowest *H/Z* value for each external pressure was set to 0 (gray background; values very close to them are marked in light gray).

| External pressue *p* [GPa] | A.f | A.af1 | A.af2 | B.f | B.af1 | B.af2 | C.f | C.af1 | C.af2 | C.af3 | Minimum-enthaply structure |
|---|---|---|---|---|---|---|---|---|---|---|---|
| | | | | | *H/Z* rel. [eV/FU] | | | | | | |
| 0 | 0.0713 | 0.0000 | 0.0029 | 0.1034 | 0.0623 | 0.0622 | 0.2019 | 0.0856 | 0.0867 | 0.0856 | **A.af1** |
| 5 | 0.0813 | 0.0097 | 0.0144 | 0.0435 | 0.0002 | 0.0000 | 0.1700 | 0.0668 | 0.0680 | 0.0668 | **B.af2** |
| 10 | 0.1710 | 0.0834 | 0.0862 | 0.0565 | 0.0005 | 0.0000 | 0.1780 | 0.0825 | 0.0842 | 0.0824 | **B.af2** |
| 20 | 0.2307 | 0.1491 | 0.1523 | 0.0763 | 0.0003 | 0.0000 | 0.2246 | 0.1323 | 0.1355 | 0.1323 | **B.af2** |
| 30 | 0.1513 | 0.0634 | 0.0644 | 0.1036 | 0.0000 | 0.0014 | 0.2485 | 0.1692 | 0.1755 | 0.1693 | **B.af1** |
| 40 | 0.2628 | 0.1377 | 0.1394 | 0.1924 | 0.0692 | 0.0710 | 0.0000 | 0.0011 | 0.0025 | 0.0011 | **C.f** |
| 50 | 0.3095 | 0.1601 | 0.1614 | 0.2456 | 0.1075 | 0.1094 | 0.0012 | 0.0002 | 0.0000 | 0.0003 | **C.af2** |
| 60 | 0.3197 | 0.2854 | 0.2854 | 0.3260 | 0.1716 | 0.1738 | 0.0023 | 0.0009 | 0.0000 | 0.0009 | **C.af2** |
| 70 | 0.3985 | 0.3207 | 0.3207 | 0.4045 | 0.2328 | 0.2349 | 0.0031 | 0.0015 | 0.0000 | 0.0015 | **C.af2** |
| 80 | 0.4751 | 0.3584 | 0.3585 | 0.4808 | 0.2903 | 0.2922 | 0.0038 | 0.0020 | 0.0000 | 0.0020 | **C.af2** |
| 90 | 0.5492 | 0.3983 | 0.3334 | 0.5547 | 0.3382 | 0.3383 | 0.0043 | 0.0025 | 0.0000 | 0.0025 | **C.af2** |
| 100 | 17.6532 | 17.5674 | 17.5700 | 0.6266 | 0.3847 | 0.3849 | 0.0048 | 0.0029 | 0.0000 | 0.0029 | **C.af2** |

**Table S2.** Relative enthalpies (*H/Z* rel.) of investigated magnetic models of chosen **RbAgF$_3$** structures optimized at DFT+U(PBEsol) level of theory under external pressures *p* from the range from 0 to 100 GPa. To facilitate data analysis data are shown as *H/Z* ratio (examined unit cells were of a different *Z* parameter values). Relative enthalpies *H/Z* rel. are given with the assumption that the lowest *H/Z* value for each external pressure was set to 0 (gray background).

| External pressue *p* [GPa] | A.f | A.af1 | A.af2 | B.f | B.af1 | B.af2 | C.f | C.af1 | C.af2 | C.af3 | Minimum-enthaply structure |
|---|---|---|---|---|---|---|---|---|---|---|---|
| | *H/Z* rel. [eV/FU] | | | *H/Z* rel. [eV/FU] | | | *H/Z* rel. [eV/FU] | | | | |
| 0   | 0.091  | 0.000  | 0.076  | 0.241 | 0.230 | 0.229 | 0.252 | 0.507 | 0.229 | 0.229 | A.af1 |
| 5   | 0.092  | 0.000  | 0.069  | 0.214 | 0.201 | 0.205 | 0.304 | 0.423 | 0.243 | 0.242 | A.af1 |
| 10  | 0.091  | 0.000  | 0.083  | 0.211 | 0.134 | 0.145 | 0.160 | 0.376 | 0.278 | 0.376 | A.af1 |
| 20  | 0.165  | 0.091  | 0.201  | 0.000 | 0.094 | 0.023 | 0.134 | 0.107 | 0.293 | 0.184 | B.f |
| 30  | 0.413  | 0.507  | 0.518  | 0.007 | 0.000 | 0.066 | 0.143 | 0.161 | 0.258 | 0.303 | B.af1 |
| 40  | 0.613  | 0.573  | 0.759  | 0.010 | 0.005 | 0.000 | 0.127 | 0.136 | 0.194 | 0.194 | B.af2 |
| 50  | 0.775  | 0.746  | 0.965  | 0.009 | 0.006 | 0.000 | 0.108 | 0.109 | 0.120 | 0.109 | B.af2 |
| 60  | 0.908  | 0.888  | 1.118  | 0.009 | 0.007 | 0.000 | 0.036 | 0.035 | 0.032 | 0.034 | B.af2 |
| 70  | 1.064  | 1.050  | 1.274  | 0.056 | 0.055 | 0.046 | 0.008 | 0.006 | 0.000 | 0.004 | C.af2 |
| 80  | 1.230  | 1.453  | 1.451  | 0.132 | 0.132 | 0.453 | 0.011 | 0.008 | 0.000 | 0.005 | C.af2 |
| 90  | 1.378  | 1.601  | 1.605  | 0.206 | 0.215 | 0.205 | 0.013 | 0.010 | 0.000 | 0.006 | C.af2 |
| 100 | 17.391 | 17.284 | 17.276 | 0.278 | 0.287 | 0.275 | 0.015 | 0.011 | 0.000 | 0.006 | C.af2 |

**Table S3.** Relative enthalpies (*H/Z* rel.) of investigated magnetic models of chosen **CsAgF$_3$** structures optimized at DFT+U(PBEsol) level of theory under external pressures ***p*** from the range from 0 to 100 GPa. To facilitate data analysis data are shown as ***H/Z*** ratio (examined unit cells were of a different ***Z*** parameter values). Relative enthalpies ***H/Z* rel.** are given with the assumption that the lowest *H/Z* value for each external pressure was set to 0 (gray background).

| External pressue *p* [GPa] | A.f | A.af1 | A.af2 | D.f | D.af1 | D.af2 | D.af3 | Minimum-enthaply structure |
|---|---|---|---|---|---|---|---|---|
| | | *H/Z* rel. [eV/FU] | | | *H/Z* rel. [eV/FU] | | | |
| 0 | 0.107 | 0.000 | 0.010 | 0.481 | 0.469 | 0.474 | 0.469 | A.af1 |
| 5 | 0.111 | 0.000 | 0.017 | 0.550 | 0.536 | 0.541 | 0.536 | A.af1 |
| 10 | 0.118 | 0.000 | 0.022 | 0.597 | 0.581 | 0.586 | 0.581 | A.af1 |
| 20 | 0.082 | 0.000 | 0.000 | 0.351 | 0.320 | 0.326 | 0.320 | A.af1 |
| 30 | 0.139 | 0.000 | 0.015 | 0.145 | 0.110 | 0.116 | 0.110 | A.af1 |
| 40 | 0.290 | 0.135 | 0.135 | 0.094 | 0.058 | 0.000 | 0.058 | D.af2 |
| 50 | 12.953 | 0.407 | 0.407 | 0.039 | 0.000 | 0.007 | 0.000 | D.af1 |
| 60 | 0.804 | 0.640 | 0.640 | 0.105 | 0.000 | 0.007 | 0.000 | D.af1 |
| 70 | 0.831 | 0.855 | 0.856 | 0.040 | 0.000 | 0.007 | 0.000 | D.af1 |
| 80 | 0.934 | 1.040 | 1.040 | 0.039 | 0.000 | 0.057 | 0.000 | D.af1 |
| 90 | 1.018 | 1.190 | 1.190 | 0.039 | 0.000 | 0.007 | 0.000 | D.af1 |
| 100 | 1.982 | 1.326 | 1.326 | 0.038 | 0.000 | 0.006 | 0.000 | D.af1 |

**Appendix S1.** CIF files of structures discussed in this study.

Ambient pressure structures.

```
## DFT(PBEsol)+U KAgF3 A p=0GPa magn.: af1
## Symmetry determined by FINDSYM v. 7.1.3
_cell_length_a     6.4221200000
_cell_length_b     8.2858800000
_cell_length_c     6.0290000000
_cell_angle_alpha 90.0000000000
_cell_angle_beta  90.0000000000
_cell_angle_gamma 90.0000000000
_cell_volume       320.8206685479

_symmetry_space_group_name_H-M "P 21/n 21/m 21/a"
_symmetry_Int_Tables_number 62
_space_group.reference_setting '062:-P 2ac 2n'
_space_group.transform_Pp_abc a,b,c;0,0,0

loop_
_space_group_symop_id
_space_group_symop_operation_xyz
1 x,y,z
2 x+1/2,-y+1/2,-z+1/2
3 -x,y+1/2,-z
4 -x+1/2,-y,z+1/2
5 -x,-y,-z
6 -x+1/2,y+1/2,z+1/2
7 x,-y+1/2,z
8 x+1/2,y,-z+1/2

loop_
_atom_site_label
_atom_site_type_symbol
_atom_site_symmetry_multiplicity
_atom_site_Wyckoff_label
_atom_site_fract_x
_atom_site_fract_y
_atom_site_fract_z
_atom_site_occupancy
_atom_site_fract_symmform
Ag1 Ag   4 a 0.00000 0.00000 0.00000 1.00000 0,0,0
K1  K    4 c 0.05972 0.25000 0.48739 1.00000 Dx,0,Dz
F1  F    4 c 0.47866 0.25000 0.57631 1.00000 Dx,0,Dz
F2  F    8 d 0.31831 0.46221 0.22290 1.00000 Dx,Dy,Dz

## DFT(PBEsol)+U KAgF3 B p=0GPa magn.: af1
## Symmetry determined by FINDSYM v. 7.1.3
_cell_length_a     3.4743000000
_cell_length_b     7.9336200000
_cell_length_c     5.7935400000
_cell_angle_alpha 90.0000000000
_cell_angle_beta  101.6136900000
_cell_angle_gamma 90.0000000000
_cell_volume       156.4224958587

_symmetry_space_group_name_H-M "P 1 21/m 1"
_symmetry_Int_Tables_number 11
_space_group.reference_setting '011:-P 2yb'
```

```
_space_group.transform_Pp_abc a,b,c;0,0,0

loop_
_space_group_symop_id
_space_group_symop_operation_xyz
1 x,y,z
2 -x,y+1/2,-z
3 -x,-y,-z
4 x,-y+1/2,z

loop_
_atom_site_label
_atom_site_type_symbol
_atom_site_symmetry_multiplicity
_atom_site_Wyckoff_label
_atom_site_fract_x
_atom_site_fract_y
_atom_site_fract_z
_atom_site_occupancy
_atom_site_fract_symmform
Ag1  Ag   2 b  0.50000   0.00000   0.00000  1.00000 0,0,0
F1   F    2 e  0.41744   0.25000   0.86894  1.00000 Dx,0,Dz
F2   F    4 f  0.80880  -0.05005   0.73826  1.00000 Dx,Dy,Dz
K1   K    2 e  0.78877   0.25000   0.50612  1.00000 Dx,0,Dz

## DFT(PBEsol)+U KAgF3 C p=0GPa magn.: af2
## Symmetry determined by FINDSYM v. 7.1.3
_cell_length_a     4.7446300000
_cell_length_b     3.7967950000
_cell_length_c    10.6613000000
_cell_angle_alpha 90.0000000000
_cell_angle_beta  90.0000000000
_cell_angle_gamma 90.0000000000
_cell_volume      192.0567890364

_symmetry_space_group_name_H-M "P 21/n 21/m 21/a"
_symmetry_Int_Tables_number 62
_space_group.reference_setting '062:-P 2ac 2n'
_space_group.transform_Pp_abc a,b,c;0,0,0

loop_
_space_group_symop_id
_space_group_symop_operation_xyz
1 x,y,z
2 x+1/2,-y+1/2,-z+1/2
3 -x,y+1/2,-z
4 -x+1/2,-y,z+1/2
5 -x,-y,-z
6 -x+1/2,y+1/2,z+1/2
7 x,-y+1/2,z
8 x+1/2,y,-z+1/2

loop_
_atom_site_label
_atom_site_type_symbol
_atom_site_symmetry_multiplicity
_atom_site_Wyckoff_label
_atom_site_fract_x
_atom_site_fract_y
_atom_site_fract_z
```

```
_atom_site_occupancy
_atom_site_fract_symmform
Ag1 Ag   4 c  0.32817  0.25000  0.43408 1.00000 Dx,0,Dz
F1  F    4 c  0.21519  0.25000  0.25660 1.00000 Dx,0,Dz
F2  F    4 c -0.00957  0.25000  0.88725 1.00000 Dx,0,Dz
F3  F    4 c  0.34994  0.25000  0.04030 1.00000 Dx,0,Dz
K1  K    4 c  0.53038  0.25000  0.82789 1.00000 Dx,0,Dz
```

## DFT(PBEsol)+U RbAgF3 A p=0GPa magn.: af1
```
## Symmetry determined by FINDSYM v. 7.1.3
_cell_length_a      6.2715100000
_cell_length_b      8.4238900000
_cell_length_c      6.3487000000
_cell_angle_alpha  90.0000000000
_cell_angle_beta   90.0000000000
_cell_angle_gamma  90.0000000000
_cell_volume        335.4050612108

_symmetry_space_group_name_H-M "P 21/n 21/m 21/a"
_symmetry_Int_Tables_number 62
_space_group.reference_setting '062:-P 2ac 2n'
_space_group.transform_Pp_abc a,b,c;0,0,0

loop_
_space_group_symop_id
_space_group_symop_operation_xyz
1 x,y,z
2 x+1/2,-y+1/2,-z+1/2
3 -x,y+1/2,-z
4 -x+1/2,-y,z+1/2
5 -x,-y,-z
6 -x+1/2,y+1/2,z+1/2
7 x,-y+1/2,z
8 x+1/2,y,-z+1/2

loop_
_atom_site_label
_atom_site_type_symbol
_atom_site_symmetry_multiplicity
_atom_site_Wyckoff_label
_atom_site_fract_x
_atom_site_fract_y
_atom_site_fract_z
_atom_site_occupancy
_atom_site_fract_symmform
Ag1 Ag   4 b  0.00000  0.00000  0.50000 1.00000 0,0,0
Rb1 Rb   4 c -0.00379  0.25000  0.00078 1.00000 Dx,0,Dz
F1  F    4 c  0.50417  0.25000  0.04188 1.00000 Dx,0,Dz
F2  F    8 d  0.26423  0.47881  0.77602 1.00000 Dx,Dy,Dz
```

## DFT(PBEsol)+U RbAgF3 B p=0GPa magn.: af2
```
## Symmetry determined by FINDSYM v. 7.1.3
_cell_length_a      7.0655900000
_cell_length_b      6.8220397223
```
S18

```
_cell_length_c     3.5165400000
_cell_angle_alpha 90.0000000000
_cell_angle_beta  90.0000000000
_cell_angle_gamma 90.0000000000
_cell_volume       169.5033314529

_symmetry_space_group_name_H-M "P 21/m 2/m 2/a"
_symmetry_Int_Tables_number 51
_space_group.reference_setting '051:-P 2a 2a'
_space_group.transform_Pp_abc a,b,c;0,0,0

loop_
_space_group_symop_id
_space_group_symop_operation_xyz
1 x,y,z
2 x+1/2,-y,-z
3 -x,y,-z
4 -x+1/2,-y,z
5 -x,-y,-z
6 -x+1/2,y,z
7 x,-y,z
8 x+1/2,y,-z

loop_
_atom_site_label
_atom_site_type_symbol
_atom_site_symmetry_multiplicity
_atom_site_Wyckoff_label
_atom_site_fract_x
_atom_site_fract_y
_atom_site_fract_z
_atom_site_occupancy
_atom_site_fract_symmform
Ag1 Ag   2 b  0.00000  0.50000  0.00000  1.00000  0,0,0
F1  F    4 g  0.00000  0.79634  0.00000  1.00000  0,Dy,0
F2  F    2 f  0.25000  0.50000  0.34959  1.00000  0,0,Dz
Rb1 Rb   2 e  0.25000  0.00000  0.49663  1.00000  0,0,Dz
```

## DFT(PBEsol)+U RbAgF3 C p=0GPa magn.: af3
```
## Symmetry determined by FINDSYM v. 7.1.3
_cell_length_a     6.3404700000
_cell_length_b     4.2261650000
_cell_length_c    13.0313500000
_cell_angle_alpha 90.0000000000
_cell_angle_beta  90.0000000000
_cell_angle_gamma 90.0000000000
_cell_volume       349.1863917678

_symmetry_space_group_name_H-M "P 21/n 21/m 21/a"
_symmetry_Int_Tables_number 62
_space_group.reference_setting '062:-P 2ac 2n'
_space_group.transform_Pp_abc a,b,c;0,0,0

loop_
```

```
_space_group_symop_id
_space_group_symop_operation_xyz
1 x,y,z
2 x+1/2,-y+1/2,-z+1/2
3 -x,y+1/2,-z
4 -x+1/2,-y,z+1/2
5 -x,-y,-z
6 -x+1/2,y+1/2,z+1/2
7 x,-y+1/2,z
8 x+1/2,y,-z+1/2

loop_
_atom_site_label
_atom_site_type_symbol
_atom_site_symmetry_multiplicity
_atom_site_Wyckoff_label
_atom_site_fract_x
_atom_site_fract_y
_atom_site_fract_z
_atom_site_occupancy
_atom_site_fract_symmform
Ag1  Ag   4  c   0.46435  0.25000   0.62494  1.00000 Dx,0,Dz
F1   F    4  c   0.22640  0.25000   0.78376  1.00000 Dx,0,Dz
F2   F    4  c   0.03833  0.25000   0.12877  1.00000 Dx,0,Dz
F3   F    4  c   0.69555  0.25000  -0.04191  1.00000 Dx,0,Dz
Rb1  Rb   4  c   0.50840  0.25000   0.14379  1.00000 Dx,0,Dz
```

## DFT(PBEsol)+U CsAgF3 A p=0GPa magn.: af1
```
## Symmetry determined by FINDSYM v. 7.1.3
_cell_length_a     6.4364900000
_cell_length_b     6.4364900000
_cell_length_c     8.5406700000
_cell_angle_alpha 90.0000000000
_cell_angle_beta  90.0000000000
_cell_angle_gamma 90.0000000000
_cell_volume      353.8263230920

_symmetry_space_group_name_H-M "I 4/m 2/c 2/m"
_symmetry_Int_Tables_number 140
_space_group.reference_setting '140:-I 4 2c'
_space_group.transform_Pp_abc a,b,c;0,0,0

loop_
_space_group_symop_id
_space_group_symop_operation_xyz
1 x,y,z
2 x,-y,-z+1/2
3 -x,y,-z+1/2
4 -x,-y,z
5 -y,-x,-z+1/2
6 -y,x,z
7 y,-x,z
8 y,x,-z+1/2
9 -x,-y,-z
```

```
10 -x,y,z+1/2
11 x,-y,z+1/2
12 x,y,-z
13 y,x,z+1/2
14 y,-x,-z
15 -y,x,-z
16 -y,-x,z+1/2
17 x+1/2,y+1/2,z+1/2
18 x+1/2,-y+1/2,-z
19 -x+1/2,y+1/2,-z
20 -x+1/2,-y+1/2,z+1/2
21 -y+1/2,-x+1/2,-z
22 -y+1/2,x+1/2,z+1/2
23 y+1/2,-x+1/2,z+1/2
24 y+1/2,x+1/2,-z
25 -x+1/2,-y+1/2,-z+1/2
26 -x+1/2,y+1/2,z
27 x+1/2,-y+1/2,z
28 x+1/2,y+1/2,-z+1/2
29 y+1/2,x+1/2,z
30 y+1/2,-x+1/2,-z+1/2
31 -y+1/2,x+1/2,-z+1/2
32 -y+1/2,-x+1/2,z

loop_
_atom_site_label
_atom_site_type_symbol
_atom_site_symmetry_multiplicity
_atom_site_Wyckoff_label
_atom_site_fract_x
_atom_site_fract_y
_atom_site_fract_z
_atom_site_occupancy
_atom_site_fract_symmform
Ag1 Ag   4 d 0.00000 0.50000 0.00000 1.00000 0,0,0
Cs1 Cs   4 a 0.00000 0.00000 0.25000 1.00000 0,0,0
F1  F    4 b 0.00000 0.50000 0.25000 1.00000 0,0,0
F2  F    8 h 0.22635 0.72635 0.00000 1.00000 Dx,Dx,0
```

## DFT(PBEsol)+U CsAgF3 D p=0GPa magn.: af1
```
## Symmetry determined by FINDSYM v. 7.1.3
_cell_length_a    7.1246700000
_cell_length_b    7.1246900000
_cell_length_c    7.1577600000
_cell_angle_alpha 65.1803200000
_cell_angle_beta  72.3921700000
_cell_angle_gamma 65.7824400000
_cell_volume      297.0031182439

_symmetry_space_group_name_H-M "P -1"
_symmetry_Int_Tables_number 2
_space_group.reference_setting '002:-P 1'
_space_group.transform_Pp_abc a,b,c;0,0,0
```

```
loop_
_space_group_symop_id
_space_group_symop_operation_xyz
1 x,y,z
2 -x,-y,-z

loop_
_atom_site_label
_atom_site_type_symbol
_atom_site_symmetry_multiplicity
_atom_site_Wyckoff_label
_atom_site_fract_x
_atom_site_fract_y
_atom_site_fract_z
_atom_site_occupancy
_atom_site_fract_symmform
Ag1 Ag   2 i 0.52980 0.72877 0.41869 1.00000 Dx,Dy,Dz
Ag2 Ag   2 i 0.01347 0.18926 0.75103 1.00000 Dx,Dy,Dz
Cs1 Cs   2 i 0.48484 0.22914 0.10565 1.00000 Dx,Dy,Dz
Cs2 Cs   2 i 0.01990 0.70090 0.72471 1.00000 Dx,Dy,Dz
F1  F    2 i 0.24952 0.00529 0.58480 1.00000 Dx,Dy,Dz
F2  F    2 i 0.42888 0.54234 0.71237 1.00000 Dx,Dy,Dz
F3  F    2 i 0.25549 0.19571 0.86567 1.00000 Dx,Dy,Dz
F4  F    2 i 0.24997 0.86999 0.28055 1.00000 Dx,Dy,Dz
F5  F    2 i 0.17134 0.65130 0.05300 1.00000 Dx,Dy,Dz
F6  F    2 i 0.80328 0.57159 0.53512 1.00000 Dx,Dy,Dz
```

Relevant elevated pressure structures.

## DFT(PBEsol)+U KAgF3 B p=5GPa magn.: af2
```
## Symmetry determined by FINDSYM v. 7.1.3
_cell_length_a    3.3590300000
_cell_length_b    7.8614100000
_cell_length_c    5.5209250000
_cell_angle_alpha 90.0000000000
_cell_angle_beta  102.7052500000
_cell_angle_gamma 90.0000000000
_cell_volume      142.2197331459

_symmetry_space_group_name_H-M "P 1 21/m 1"
_symmetry_Int_Tables_number 11
_space_group.reference_setting '011:-P 2yb'
_space_group.transform_Pp_abc a,b,c;0,0,0

loop_
_space_group_symop_id
_space_group_symop_operation_xyz
1 x,y,z
2 -x,y+1/2,-z
3 -x,-y,-z
4 x,-y+1/2,z
```

```
loop_
_atom_site_label
_atom_site_type_symbol
_atom_site_symmetry_multiplicity
_atom_site_Wyckoff_label
_atom_site_fract_x
_atom_site_fract_y
_atom_site_fract_z
_atom_site_occupancy
_atom_site_fract_symmform
Ag1 Ag   2 b 0.50000  0.00000  0.00000 1.00000 0,0,0
F1  F    2 e 0.41157  0.25000  0.86918 1.00000 Dx,0,Dz
F2  F    4 f 0.81840 -0.04992  0.72924 1.00000 Dx,Dy,Dz
K1  K    2 e 0.77639  0.25000  0.50214 1.00000 Dx,0,Dz
```

## DFT(PBEsol)+U KAgF3 B p=30GPa magn.: af1

```
## Symmetry determined by FINDSYM v. 7.1.3
_cell_length_a    3.7092900000
_cell_length_b    8.0746900000
_cell_length_c    3.9109000000
_cell_angle_alpha 90.0000000000
_cell_angle_beta  112.8647100000
_cell_angle_gamma 90.0000000000
_cell_volume      107.9327651741

_symmetry_space_group_name_H-M "P 1 21/m 1"
_symmetry_Int_Tables_number 11
_space_group.reference_setting '011:-P 2yb'
_space_group.transform_Pp_abc a,b,c;0,0,0

loop_
_space_group_symop_id
_space_group_symop_operation_xyz
1 x,y,z
2 -x,y+1/2,-z
3 -x,-y,-z
4 x,-y+1/2,z

loop_
_atom_site_label
_atom_site_type_symbol
_atom_site_symmetry_multiplicity
_atom_site_Wyckoff_label
_atom_site_fract_x
_atom_site_fract_y
_atom_site_fract_z
_atom_site_occupancy
_atom_site_fract_symmform
Ag1 Ag   2 b 0.50000  0.00000  0.00000 1.00000 0,0,0
F1  F    2 e 0.48277  0.25000  0.03463 1.00000 Dx,0,Dz
F2  F    4 f 0.18807 -0.07044  0.30913 1.00000 Dx,Dy,Dz
K1  K    2 e 0.14218  0.25000  0.40020 1.00000 Dx,0,Dz
```

## DFT(PBEsol)+U KAgF3 C p=50GPa magn.: af1

```
## Symmetry determined by FINDSYM v. 7.1.3
_cell_length_a    4.7446300000
_cell_length_b    3.7967950000
_cell_length_c    10.6613000000
_cell_angle_alpha 90.0000000000
_cell_angle_beta  90.0000000000
_cell_angle_gamma 90.0000000000
_cell_volume      192.0567890364

_symmetry_space_group_name_H-M "P 21/n 21/m 21/a"
_symmetry_Int_Tables_number 62
_space_group.reference_setting '062:-P 2ac 2n'
_space_group.transform_Pp_abc a,b,c;0,0,0

loop_
_space_group_symop_id
_space_group_symop_operation_xyz
1 x,y,z
2 x+1/2,-y+1/2,-z+1/2
3 -x,y+1/2,-z
4 -x+1/2,-y,z+1/2
5 -x,-y,-z
6 -x+1/2,y+1/2,z+1/2
7 x,-y+1/2,z
8 x+1/2,y,-z+1/2

loop_
_atom_site_label
_atom_site_type_symbol
_atom_site_symmetry_multiplicity
_atom_site_Wyckoff_label
_atom_site_fract_x
_atom_site_fract_y
_atom_site_fract_z
_atom_site_occupancy
_atom_site_fract_symmform
Ag1 Ag   4 c 0.32817  0.25000 0.43408 1.00000 Dx,0,Dz
F1  F    4 c 0.21519  0.25000 0.25660 1.00000 Dx,0,Dz
F2  F    4 c -0.00957 0.25000 0.88725 1.00000 Dx,0,Dz
F3  F    4 c 0.34994  0.25000 0.04030 1.00000 Dx,0,Dz
K1  K    4 c 0.53038  0.25000 0.82789 1.00000 Dx,0,Dz

## **DFT(PBEsol)+U KAgF3 C p=100GPa magn.: af2**
## Symmetry determined by FINDSYM v. 7.1.3
_cell_length_a    4.4594200000
_cell_length_b    3.6803500000
_cell_length_c    10.1469100000
_cell_angle_alpha 90.0000000000
_cell_angle_beta  90.0000000000
_cell_angle_gamma 90.0000000000
_cell_volume      166.5333841500

_symmetry_space_group_name_H-M "P 21/n 21/m 21/a"
_symmetry_Int_Tables_number 62
```

```
_space_group.reference_setting '062:-P 2ac 2n'
_space_group.transform_Pp_abc a,b,c;0,0,0

loop_
_space_group_symop_id
_space_group_symop_operation_xyz
1 x,y,z
2 x+1/2,-y+1/2,-z+1/2
3 -x,y+1/2,-z
4 -x+1/2,-y,z+1/2
5 -x,-y,-z
6 -x+1/2,y+1/2,z+1/2
7 x,-y+1/2,z
8 x+1/2,y,-z+1/2

loop_
_atom_site_label
_atom_site_type_symbol
_atom_site_symmetry_multiplicity
_atom_site_Wyckoff_label
_atom_site_fract_x
_atom_site_fract_y
_atom_site_fract_z
_atom_site_occupancy
_atom_site_fract_symmform
Ag1  Ag   4 c  0.32705   0.25000  0.42932  1.00000  Dx,0,Dz
F1   F    4 c  0.20295   0.25000  0.24616  1.00000  Dx,0,Dz
F2   F    4 c  -0.01127  0.25000  0.88628  1.00000  Dx,0,Dz
F3   F    4 c  0.34966   0.25000  0.04106  1.00000  Dx,0,Dz
K1   K    4 c  0.52316   0.25000  0.82855  1.00000  Dx,0,Dz

## DFT(PBEsol)+U RbAgF3 A p=10GPa magn.: af1
## Symmetry determined by FINDSYM v. 7.1.3
_cell_length_a     5.8130400000
_cell_length_b     8.0600000000
_cell_length_c     6.2073000000
_cell_angle_alpha  90.0000000000
_cell_angle_beta   90.0000000000
_cell_angle_gamma  90.0000000000
_cell_volume       290.8312625275

_symmetry_space_group_name_H-M "P 21/n 21/m 21/a"
_symmetry_Int_Tables_number 62
_space_group.reference_setting '062:-P 2ac 2n'
_space_group.transform_Pp_abc a,b,c;0,0,0

loop_
_space_group_symop_id
_space_group_symop_operation_xyz
1 x,y,z
2 x+1/2,-y+1/2,-z+1/2
3 -x,y+1/2,-z
4 -x+1/2,-y,z+1/2
5 -x,-y,-z
```

```
6  -x+1/2,y+1/2,z+1/2
7  x,-y+1/2,z
8  x+1/2,y,-z+1/2

loop_
_atom_site_label
_atom_site_type_symbol
_atom_site_symmetry_multiplicity
_atom_site_Wyckoff_label
_atom_site_fract_x
_atom_site_fract_y
_atom_site_fract_z
_atom_site_occupancy
_atom_site_fract_symmform
Ag1 Ag   4 b  0.00000   0.00000  0.50000 1.00000 0,0,0
Rb1 Rb   4 c -0.00758   0.25000  0.00184 1.00000 Dx,0,Dz
F1  F    4 c  0.49692   0.25000  0.06960 1.00000 Dx,0,Dz
F2  F    8 d  0.25646   0.46633  0.76972 1.00000 Dx,Dy,Dz
```

## **DFT(PBEsol)+U RbAgF3 B p=20GPa magn.: f**
```
## Symmetry determined by FINDSYM v. 7.1.3
_cell_length_a     2.9471300000
_cell_length_b     7.1260500000
_cell_length_c     5.9299100000
_cell_angle_alpha  90.0000000000
_cell_angle_beta   98.5780500000
_cell_angle_gamma  90.0000000000
_cell_volume       123.1432721894

_symmetry_space_group_name_H-M "P 1 21/m 1"
_symmetry_Int_Tables_number 11
_space_group.reference_setting '011:-P 2yb'
_space_group.transform_Pp_abc a,b,c;0,0,0

loop_
_space_group_symop_id
_space_group_symop_operation_xyz
1 x,y,z
2 -x,y+1/2,-z
3 -x,-y,-z
4 x,-y+1/2,z

loop_
_atom_site_label
_atom_site_type_symbol
_atom_site_symmetry_multiplicity
_atom_site_Wyckoff_label
_atom_site_fract_x
_atom_site_fract_y
_atom_site_fract_z
_atom_site_occupancy
_atom_site_fract_symmform
Ag1 Ag   2 b  0.50000 0.00000 0.00000 1.00000 0,0,0
F1  F    4 f  0.05533 0.45781 0.71134 1.00000 Dx,Dy,Dz
```

```
F2   F    2 e 0.87310 0.25000 0.04325 1.00000 Dx,0,Dz
Rb1  Rb   2 e 0.48317 0.25000 0.45212 1.00000 Dx,0,Dz
```

## DFT(PBEsol)+U RbAgF3 B p=60GPa magn.: af2
```
## Symmetry determined by FINDSYM v. 7.1.3
_cell_length_a    2.7787350000
_cell_length_b    6.6106700000
_cell_length_c    5.5542200000
_cell_angle_alpha 90.0000000000
_cell_angle_beta  96.1940800000
_cell_angle_gamma 90.0000000000
_cell_volume      101.4315115985

_symmetry_space_group_name_H-M "P 1 21/m 1"
_symmetry_Int_Tables_number 11
_space_group.reference_setting '011:-P 2yb'
_space_group.transform_Pp_abc a,b,c;0,0,0

loop_
_space_group_symop_id
_space_group_symop_operation_xyz
1 x,y,z
2 -x,y+1/2,-z
3 -x,-y,-z
4 x,-y+1/2,z

loop_
_atom_site_label
_atom_site_type_symbol
_atom_site_symmetry_multiplicity
_atom_site_Wyckoff_label
_atom_site_fract_x
_atom_site_fract_y
_atom_site_fract_z
_atom_site_occupancy
_atom_site_fract_symmform
Ag1  Ag   2 a 0.00000 0.00000  0.00000 1.00000 0,0,0
F1   F    4 f 0.45741 0.05008  0.29301 1.00000 Dx,Dy,Dz
F2   F    2 e 0.57165 0.25000 -0.04996 1.00000 Dx,0,Dz
Rb1  Rb   2 e 0.00975 0.25000  0.55795 1.00000 Dx,0,Dz
```

## DFT(PBEsol)+U RbAgF3 C p=70GPa magn.: af2
```
## Symmetry determined by FINDSYM v. 7.1.3
_cell_length_a    4.6299000000
_cell_length_b    3.9309650000
_cell_length_c    10.5272900000
_cell_angle_alpha 90.0000000000
_cell_angle_beta  90.0000000000
_cell_angle_gamma 90.0000000000
_cell_volume      191.5964132755

_symmetry_space_group_name_H-M "P 21/n 21/m 21/a"
_symmetry_Int_Tables_number 62
_space_group.reference_setting '062:-P 2ac 2n'
```

```
_space_group.transform_Pp_abc a,b,c;0,0,0

loop_
_space_group_symop_id
_space_group_symop_operation_xyz
1 x,y,z
2 x+1/2,-y+1/2,-z+1/2
3 -x,y+1/2,-z
4 -x+1/2,-y,z+1/2
5 -x,-y,-z
6 -x+1/2,y+1/2,z+1/2
7 x,-y+1/2,z
8 x+1/2,y,-z+1/2

loop_
_atom_site_label
_atom_site_type_symbol
_atom_site_symmetry_multiplicity
_atom_site_Wyckoff_label
_atom_site_fract_x
_atom_site_fract_y
_atom_site_fract_z
_atom_site_occupancy
_atom_site_fract_symmform
Ag1 Ag   4 c  0.33196  0.25000  0.57782 1.00000 Dx,0,Dz
F1  F    4 c  0.19660  0.25000  0.75597 1.00000 Dx,0,Dz
F2  F    4 c -0.00322  0.25000  0.09924 1.00000 Dx,0,Dz
F3  F    4 c  0.36081  0.25000 -0.04234 1.00000 Dx,0,Dz
Rb1 Rb   4 c  0.52863  0.25000  0.17216 1.00000 Dx,0,Dz
```

## DFT(PBEsol)+U RbAgF3 C p=100GPa magn.: af2
```
## Symmetry determined by FINDSYM v. 7.1.3
_cell_length_a     4.5189100000
_cell_length_b     3.8313650000
_cell_length_c    10.2600900000
_cell_angle_alpha 90.0000000000
_cell_angle_beta  90.0000000000
_cell_angle_gamma 90.0000000000
_cell_volume     177.6390286841

_symmetry_space_group_name_H-M "P 21/n 21/m 21/a"
_symmetry_Int_Tables_number 62
_space_group.reference_setting '062:-P 2ac 2n'
_space_group.transform_Pp_abc a,b,c;0,0,0

loop_
_space_group_symop_id
_space_group_symop_operation_xyz
1 x,y,z
2 x+1/2,-y+1/2,-z+1/2
3 -x,y+1/2,-z
4 -x+1/2,-y,z+1/2
5 -x,-y,-z
6 -x+1/2,y+1/2,z+1/2
7 x,-y+1/2,z
```

```
8 x+1/2,y,-z+1/2

loop_
_atom_site_label
_atom_site_type_symbol
_atom_site_symmetry_multiplicity
_atom_site_Wyckoff_label
_atom_site_fract_x
_atom_site_fract_y
_atom_site_fract_z
_atom_site_occupancy
_atom_site_fract_symmform
Ag1 Ag   4 c  0.67295  0.25000  0.57917  1.00000 Dx,0,Dz
F1  F    4 c  0.81101  0.25000  0.76080  1.00000 Dx,0,Dz
F2  F    4 c  0.00589  0.25000  0.09954  1.00000 Dx,0,Dz
F3  F    4 c  0.64263  0.25000 -0.04160  1.00000 Dx,0,Dz
Rb1 Rb   4 c  0.47289  0.25000  0.17257  1.00000 Dx,0,Dz
```

## DFT(PBEsol)+U CsAgF3 A p=30GPa magn.: af1
```
## Symmetry determined by FINDSYM v. 7.1.3
_cell_length_a    5.7832500000
_cell_length_b    5.7832500000
_cell_length_c    7.9785700000
_cell_angle_alpha 90.0000000000
_cell_angle_beta  90.0000000000
_cell_angle_gamma 90.0000000000
_cell_volume      266.8510971365

_symmetry_space_group_name_H-M "I 4/m 2/c 2/m"
_symmetry_Int_Tables_number 140
_space_group.reference_setting '140:-I 4 2c'
_space_group.transform_Pp_abc a,b,c;0,0,0

loop_
_space_group_symop_id
_space_group_symop_operation_xyz
1 x,y,z
2 x,-y,-z+1/2
3 -x,y,-z+1/2
4 -x,-y,z
5 -y,-x,-z+1/2
6 -y,x,z
7 y,-x,z
8 y,x,-z+1/2
9 -x,-y,-z
10 -x,y,z+1/2
11 x,-y,z+1/2
12 x,y,-z
13 y,x,z+1/2
14 y,-x,-z
15 -y,x,-z
16 -y,-x,z+1/2
17 x+1/2,y+1/2,z+1/2
18 x+1/2,-y+1/2,-z
```

```
19 -x+1/2,y+1/2,-z
20 -x+1/2,-y+1/2,z+1/2
21 -y+1/2,-x+1/2,-z
22 -y+1/2,x+1/2,z+1/2
23 y+1/2,-x+1/2,z+1/2
24 y+1/2,x+1/2,-z
25 -x+1/2,-y+1/2,-z+1/2
26 -x+1/2,y+1/2,z
27 x+1/2,-y+1/2,z
28 x+1/2,y+1/2,-z+1/2
29 y+1/2,x+1/2,z
30 y+1/2,-x+1/2,-z+1/2
31 -y+1/2,x+1/2,-z+1/2
32 -y+1/2,-x+1/2,z

loop_
_atom_site_label
_atom_site_type_symbol
_atom_site_symmetry_multiplicity
_atom_site_Wyckoff_label
_atom_site_fract_x
_atom_site_fract_y
_atom_site_fract_z
_atom_site_occupancy
_atom_site_fract_symmform
Ag1 Ag   4 d 0.00000 0.50000 0.00000 1.00000 0,0,0
Cs1 Cs   4 a 0.00000 0.00000 0.25000 1.00000 0,0,0
F1  F    4 b 0.00000 0.50000 0.25000 1.00000 0,0,0
F2  F    8 h 0.24153 0.74153 0.00000 1.00000 Dx,Dx,0
```

## DFT(PBEsol)+U CsAgF3 D p=40GPa magn.: af2

```
## Symmetry determined by FINDSYM v. 7.1.3
_cell_length_a     6.3187000000
_cell_length_b     6.7179000000
_cell_length_c     6.9605000000
_cell_angle_alpha 69.7356700000
_cell_angle_beta  66.6788600000
_cell_angle_gamma 62.5709600000
_cell_volume       235.8968168451

_symmetry_space_group_name_H-M "P -1"
_symmetry_Int_Tables_number 2
_space_group.reference_setting '002:-P 1'
_space_group.transform_Pp_abc a,b,c;0,0,0

loop_
_space_group_symop_id
_space_group_symop_operation_xyz
1 x,y,z
2 -x,-y,-z

loop_
_atom_site_label
_atom_site_type_symbol
```

```
_atom_site_symmetry_multiplicity
_atom_site_Wyckoff_label
_atom_site_fract_x
_atom_site_fract_y
_atom_site_fract_z
_atom_site_occupancy
_atom_site_fract_symmform
Ag1 Ag   2 i  0.73012  0.41565 0.55351 1.00000 Dx,Dy,Dz
Ag2 Ag   2 i  0.15380  0.76846 0.00901 1.00000 Dx,Dy,Dz
Cs1 Cs   2 i  0.22489  0.11390 0.46526 1.00000 Dx,Dy,Dz
Cs2 Cs   2 i  0.69783  0.71489 0.02313 1.00000 Dx,Dy,Dz
F1  F    2 i -0.00026  0.57285 0.27550 1.00000 Dx,Dy,Dz
F2  F    2 i  0.53775  0.72944 0.41919 1.00000 Dx,Dy,Dz
F3  F    2 i  0.19150  0.89963 0.22133 1.00000 Dx,Dy,Dz
F4  F    2 i  0.85599  0.28244 0.27246 1.00000 Dx,Dy,Dz
F5  F    2 i  0.68151  0.01709 0.17974 1.00000 Dx,Dy,Dz
F6  F    2 i  0.55165  0.56664 0.80940 1.00000 Dx,Dy,Dz
```

## DFT(PBEsol)+U CsAgF3 D p=100GPa magn.: af1

```
## Symmetry determined by FINDSYM v. 7.1.3
_cell_length_a    5.8549900000
_cell_length_b    6.3062628381
_cell_length_c    6.6311500000
_cell_angle_alpha 89.9276707364
_cell_angle_beta  66.6953300000
_cell_angle_gamma 63.2763018059
_cell_volume      196.1359887276

_symmetry_space_group_name_H-M "P -1"
_symmetry_Int_Tables_number 2
_space_group.reference_setting '002:-P 1'
_space_group.transform_Pp_abc a,b,c;0,0,0

loop_
_space_group_symop_id
_space_group_symop_operation_xyz
1 x,y,z
2 -x,-y,-z

loop_
_atom_site_label
_atom_site_type_symbol
_atom_site_symmetry_multiplicity
_atom_site_Wyckoff_label
_atom_site_fract_x
_atom_site_fract_y
_atom_site_fract_z
_atom_site_occupancy
_atom_site_fract_symmform
Ag1 Ag   2 i  0.36451  0.41208  0.43526 1.00000 Dx,Dy,Dz
Ag2 Ag   2 i  0.58501  0.77000 -0.01009 1.00000 Dx,Dy,Dz
Cs1 Cs   2 i  0.16540  0.11352  0.53454 1.00000 Dx,Dy,Dz
Cs2 Cs   2 i  0.09021  0.71164 -0.02247 1.00000 Dx,Dy,Dz
F1  F    2 i -0.06216  0.56639  0.71663 1.00000 Dx,Dy,Dz
```

```
F2   F     2 i 0.22236  0.73600 0.58451 1.00000 Dx,Dy,Dz
F3   F     2 i 0.39420 -0.08290 0.78804 1.00000 Dx,Dy,Dz
F4   F     2 i 0.36971  0.28846 0.71994 1.00000 Dx,Dy,Dz
F5   F     2 i 0.80764  0.01122 0.81890 1.00000 Dx,Dy,Dz
F6   F     2 i 0.37246  0.57903 0.18915 1.00000 Dx,Dy,Dz
```